# Anisotropic fluid star model in isotropic coordinates

**Neeraj Pant[1], Narendra Pradhan[2], Manuel Malaver[3]**

[1]Mathematics Department, National Defence Academy, Khadakwasla, Pune-411023, India
[2]Physics Department, National Defence Academy, Khadakwasla, Pune-411023, India
[3]Department of Basic Sciences, Maritime University of the Caribbean, Catia la Mar, Venezuela

**Email address:**
neeraj.pant@yahoo.com (N. Pant), npradhan20569@gmail.com (N. Pradhan), mmf.umc@gmail.com (M. Malaver)

**Abstract:** We present a spherically symmetric solution of the general relativistic field equations in isotropic coordinates for anisotropic neutral fluid, compatible with a super dense star modeling by considering a specific choice of anisotropy factor Δ that includes a positive constant "α" defined as anisotropy parameter, which varies the relation between the radial and tangential pressure. Further, we have constructed a super-dense star model with all degree of suitability. We have found that the maximum mass decreases with the increase of anisotropy parameter (α). The robustness of our result is that it matches with the recent discoveries.

## 1.Introduction

Of course no astrophysical object is composed of purely perfect fluid. In the formulism of realistic model of super dense stars, it is also important to include the pressure anisotropy. Since the theoretical investigations of Ruderman (1972) about more realistic stellar models show that the stellar matter may be anisotropic at least in certain very high density ranges $\rho > 10^{15}$ g/cc, where the nuclear interactions must be treated relativistically. According to these views in such massive stellar objects the radial pressure may not be equal to the tangential one. Bowers and Liang (1974) have extensively investigated the possible importance of locally anisotropic equations of state for relativistic fluid spheres by generalizing the equations of hydrostatic equilibrium to include the effect of local anisotropy. Their study shows that anisotropy may have non-negligible effects on such parameters as maximum equilibrium mass and surface red shift. Due to the uncertainty in the behavior of matter in highly compact stars like normal matter neutron stars or self-bound strange quark matter (SQM) stars insight into the structure of such astrophysical objects can be obtained by reference to applicable analytic solutions of Einstein's gravitational field equations. Even though a few spherically symmetric analytical solutions (in both curvature and isotropic coordinates) are relevant in the modeling of compact stars (Delgaty and Lake 1998) researchers have been continuously proposing different models of astrophysical objects immense gravity by considering the distinct nature of matter or radiation (energy-momentum tensor) present in them. Analytical solutions of the field equations for various neutral/charged static spherically symmetric configurations for anisotropic pressure compatible to compact stellar modeling have been obtained in numerous works. Some of them include (Herrera and Leon 1985, Herrera and Santos 1997, Herrera et al. 2001, Mak and Harko 2002, Mak et al. 2002, Mak and Harko 2003, Chaisi and Maharaj 2006, Sharma and Maharaj 2007, Herrera et al. 2008, Maharaj and Takisa 2012, Thirukkanesh and Ragel 2012, Maurya and Gupta 2012; 2013; 2014, Takisa and Maharaj 2013a; 2013b, Maharaj et al. 2014, Malaver 2013;2014, Pant et al 2014a;2014b, Maurya and Gupta,2013;2014).

## 2.Conditions for Well behaved Solution

For well behaved nature of the solution in isotropic coordinates, the following conditions should be satisfied:
(i)The solution should be free from physical and geometrical singularities i.e. finite and positive values of central pressure, central density and non zero positive values of $e^{\omega}$ and $e^{\upsilon}$.

(ii) The solution should have positive and monotonically decreasing expressions for radial pressure, transversal pressure and density with the increase of *r*.

(iii) The solution should have positive value of ratio trace of energy stress tensor to energy Density, $Q = (P_r + 2P_t)/\rho$ and less than 1(weak energy condition) and less than 1/3 (strong energy condition) throughout within the star, monotonically decreasing as well, Esculpi et al [2007].

(iv) The casualty condition should be obeyed i.e. velocity of sound should be less than that of light throughout the model. In addition to the above the velocity of sound should be decreasing towards the surface i.e. $\frac{d}{dr}\left(\frac{dp}{d\rho}\right) < 0$ or $\left(\frac{d^2p}{d\rho^2}\right) > 0$ for $0 \le r \le r_b$ i.e. the velocity of sound is increasing with the increase of density. In this context it is worth mentioning that the equation of state at ultra-high distribution, has the property that the sound speed is decreasing outwards, Canuto [1975].

$\frac{p}{\rho} \le \frac{dp}{d\rho}$, everywhere within the ball. $\gamma = \frac{d\log_e P}{d\log_e \rho} = \frac{\rho}{p}\frac{dp}{d\rho} \Rightarrow \frac{dp}{d\rho} = \gamma \frac{p}{\rho}$, for realistic matter $\gamma \ge 1$. The stiffness parameter should increase from its lowest value 4/3 from the center to infinity.

(v) The red shift z should be positive, finite and monotonically decreasing in nature with the increase of *r*.

(vi) The anisotropy factor should be zero at the center and be increasing towards the surface.

## 3. Field Equations in Isotropic Coordinates

We consider the static and spherically symmetric metric in isotropic co-ordinates

$$ds^2 = -e^{\omega}\left[\,dr^2 + r^2(d\theta^2 + \sin^2\theta d\phi^2)\,\right] + e^{\upsilon}dt^2 \quad (1)$$

Where, $\omega$ and $\upsilon$ are functions of $r$. Einstein's field equations of gravitation for a non empty space-time are

$$R_j^i - \frac{1}{2}R\delta_j^i = -T_j^i = -[(p_\perp + \rho)v^i v_j - p_\perp \delta_j^i + (p_r - p_\perp)\chi_j \chi^i] \quad (2)$$

Where $R_j^i$ is Ricci tensor, $T_j^i$ is energy-momentum tensor, $R$ is the scalar curvature, $p_r$ denotes the radial pressure, $p_\perp$ is the transversal pressure, $\rho$ the density distribution, $\chi^i$ is the unit spacelike vector in the radial direction and $v_i$ the velocity vector, satisfying the relation

$$g_{ij}v^i v^j = 1 \quad (3)$$

Since the field is static, we have

$$v^1 = v^2 = v^3 = 0 \text{ and } v^4 = \frac{1}{\sqrt{g_{44}}} \quad (4)$$

Thus we find that for the metric (1) under these conditions and for matter distributions with anisotropic pressure the field equation (2) reduces to the following.

$$p_r = e^{-\omega}\left[\frac{(\omega')^2}{4} + \frac{\omega'}{r} + \frac{\omega'\upsilon'}{2} + \frac{\upsilon'}{r}\right] \quad (5)$$

$$p_\perp = e^{-\omega}\left[\frac{\omega''}{2} + \frac{\upsilon''}{2} + \frac{(\upsilon')^2}{4} + \frac{\omega'}{2r} + \frac{\upsilon'}{2r}\right] \quad (6)$$

$$\rho = -e^{-\omega}\left[\,\omega'' + \frac{(\omega')^2}{4} + \frac{2\omega'}{r}\,\right] \quad (7)$$

Where, prime $(')$ denotes differentiation with respect to $r$. From equations (5) and (6) we obtain following differential equation in $\omega$ and $\upsilon$.

$$e^{-\omega}\left(\frac{\omega''}{2} + \frac{\upsilon''}{2} + \frac{(\upsilon')^2}{4} - \frac{(\omega')^2}{4} - \frac{\omega'\upsilon'}{2} - \frac{\omega'}{2r} - \frac{\upsilon'}{2r}\right) - 2\frac{q^2}{r^4} = (p_\perp - p_r) \quad (8)$$

Our task is to explore the solutions of equation (8) and obtain the fluid parameters $p_r$, $p_\perp$ and $\rho$ from equation (5), (6) and (7) respectively. To solve the above equation we consider a seed solution as Pant et al [2014b].

We also take,

$$(p_\perp - p_r) = \Delta = \frac{2.\alpha.C^2 r^2 (1+Cr^2)^{\frac{-12}{7}}}{B^2} \quad (9)$$

Where $\Delta$ is the anisotropy factor whose value is zero at the center and increases towards the boundary and $\alpha$ is a positive constant defined as anisotropy parameter for Esculpi et al [2007].

## 4. Boundary Conditions in Isotropic Coordinates

For exploring the boundary conditions, we use the principle that the metric coefficients $g_{ij}$ and their first derivatives $g_{ij,k}$ in interior solution (I) as well as in exterior solution (E) are continuous upto and on the boundary B. The continuity of metric coefficients $g_{ij}$ of I and B on the boundary is known first fundamental form. The continuity of derivatives of metric coefficients $g_{ij}$ of I and B on the boundary is known second fundamental form.

The exterior field of a spherically symmetric static charged fluid distribution is described by Schwarzschild metric as

$$ds^2 = \left(1 - \frac{2GM}{c^2 R}\right)c^2 dt^2 - \left(1 - \frac{2GM}{c^2 R}\right)^{-1} dR^2 - R^2 d\theta^2 - R^2 \sin^2\theta d\phi^2 \quad (10)$$

Where M is the mass of the ball as determined by the external observer and R is the radial coordinate of the exterior region.

Since Schwarzschild metric (10) is considered as the exterior solution, thus we shall arrive at the following conclusions by matching first and second fundamental forms:

$$e^{\upsilon_b} = [1 - 2\frac{GM}{c^2 R_b}] \quad (11)$$

$$R_b = r_b \cdot e^{\frac{\omega_b}{2}} \text{ and } p_{r(r=rb)} = 0 \quad (12)$$

$$\frac{1}{2}\left(\omega' + \frac{2}{r}\right)_b r_b = \left(1 - 2\frac{GM}{c^2 R_b}\right)^{1/2} \quad (13)$$

$$\frac{1}{2}(\upsilon')_b r_b = \frac{GM}{c^2 R_b}\left(1 - 2\frac{GM}{c^2 R_b}\right)^{-1/2} \quad (14)$$

Equations (11) to (14) are four conditions, known as boundary conditions in isotropic coordinates. Moreover, (12) and (14) are equivalent to zero pressure of the interior solution on the boundary.

## 5. A New Class of Solution

The equation (8) is solved by assuming the seed solution as Pant et al [2014b]. Thus we have,

$$e^{\omega/2} = B(1 + Cr^2)^{-\frac{1}{7}}, x = Cr^2, y = \frac{d\upsilon}{dx} \text{ and}$$

$$(p_\perp - p_r) = \Delta = \frac{2.\alpha.C^2 r^2 (1 + Cr^2)^{\frac{-12}{7}}}{B^2} \quad (15)$$

On substituting the above in eqn (8), we get the following Riccati differential equation in $y$,

$$\frac{dy}{dx} + \frac{2}{7}\frac{1}{(1+x)}y + \frac{1}{2}y^2 = -\frac{12}{49}\frac{1}{(1+x)^2} + \frac{\alpha}{(1+x)^2} \quad (16)$$

Which yields the following solution,

$$e^{\frac{\upsilon}{2}} = \frac{\left\{1 + A\left(1 + Cr^2\right)^{\frac{2S}{14}}\right\}\left(1 + Cr^2\right)^{\frac{5-S}{14}}}{B^2} \quad (17)$$

Where A, B, C and K are arbitrary constants and

$$S = \sqrt{98\alpha + 1} \quad (18)$$

The expressions for density and radial pressure and tangential pressure are given by

$$\rho = \frac{C}{49B^2 f^{24}}\left(84 + 24Cr^2\right) \quad (19)$$

$$p_r = \frac{C}{49B^2 f^{24}\{1 + Af^{2S}\}}\left[\begin{array}{l}Cr^2[f^{2S}A(-16-4S) + (-16+4S)] \\ + f^{14+2S}A(42+14S) + f^{14}(42-14S)\end{array}\right] \quad (20)$$

$$p_\perp = \frac{C}{49B^2 f^{24}\{1 + Af^{2S}\}}\left[\begin{array}{l}Cr^2[f^{2S}A(98\alpha - 16 - 4S) + (98\alpha - 16 + 4S)] \\ + f^{14+2S}A(42+14S) + f^{14}(42-14S)\end{array}\right]$$

$$Where, f = (1 + Cr^2)^{\frac{1}{14}} \quad (21)$$

The trace of stress tensor to energy density ratio is given by $Q = \frac{p_r + 2p_\perp}{\rho}$

## 6. Properties of the New Solution

The central values of pressure and density are given by

$$(p_r)_{r=0} = (p_\perp)_{r=0} = \frac{2C}{7B^2(1+A)}[3(1+A) + S(A-1)] \quad (22)$$

$$(\rho)_{r=0} = \frac{12C}{7B^2} \quad (23)$$

The central values of pressure and density will be non zero positive definite, if the following conditions will be satisfied.

$$A > (S-3)/(S+3), C > 0 \quad (24)$$

Subjecting the condition that positive value of ratio of pressure-density and less than 1 at the centre i.e. $\frac{p_0}{\rho_0} \leq 1$ which leads to the following inequality,

$$\{\frac{p_r}{\rho}\}_{r=0} = (\frac{p_\perp}{\rho})_{r=0} = \frac{3(1+A) + SA - S}{6(1+A)} = \frac{1}{2} - \frac{S(1-A)}{6(1+A)} \leq 1 \quad (25)$$

All the values of A which satisfy equation (24), will also lead to the condition $\frac{p_0}{\rho_0} \leq 1$.

Differentiating (20) with respect to r, we get

$$\frac{dp_r}{dr} = \frac{C^2 r}{343B^2\{1 + Af^{2S}\}^2 f^{38}}\left[\begin{array}{l}-10.Cr^2[A^2 f^{4S}(26+10S) + (26-10S) \\ + 2Af^{2S}(26-2S^2)] \\ + 14[f^{4S} A^2\{-46-14S\} + (-46+14S) \\ + f^{2S} 2A(-46+2S^2)]\end{array}\right] \quad 26)$$

$$\frac{dp_\perp}{dr} = \frac{C^2 r}{343B^2\{1 + Af^{2S}\}^2 f^{38}}\left[\begin{array}{l}-10.Cr^2[A^2 f^{4S}(98\alpha + 26 + 10S) + (98\alpha + 26 - 10S) \\ + 2Af^{2S}(98\alpha + 26 - 2S^2)] \\ + 14[f^{4S} A^2\{98\alpha - 46 - 14S\} + (98\alpha - 46 + 14S) \\ + f^{2S} 2A(98\alpha - 46 + 2S^2)]\end{array}\right] \quad (27)$$

Thus it is found that extrema of $p_r$ and $p_\perp$ occur at the centre i.e

$$p' = 0 \Rightarrow r = 0 \text{ and } (p'')_{r=0} = -ve \quad (28)$$

Thus the expression of right hand side of equation (28) is negative for all values of A satisfying condition (24), showing thereby that the radial pressure and the transversal pressure are maximum at the centre and monotonically decreasing.

Now differentiating equation (19) with respect to r we get

$$\frac{d\rho}{dr} = \frac{C^2 r}{343B^2 f^{38}}\left[-1680 - 240Cr^2\right] \quad (29)$$

Thus the extrema of ρ occur at the centre if

$$\rho' = 0 \Rightarrow r = 0$$

$$(\rho'')_{r=0} = \frac{-1680C^2}{343B^2} \quad (30)$$

Thus, the expression of right hand side of (30) is negative showing thereby that the density ρ is maximum at the centre and monotonically decreasing.

The square of adiabatic sound speed at the centre, $\left(\frac{dp_r}{d\rho}\right)_{r=0}$, is given by

$$\left(\frac{dp_r}{d\rho}\right)_{r=0} = \frac{\left(46.(1+A)^2 - 14S - 4AS^2 + 14SA^2\right)}{120(1+A)^2} \quad <1\,\text{and positive} \quad (31)$$

The causality condition is obeyed at the centre for all values of constants satisfying condition(24).

Due to cumbersome expressions the trend of pressure-density ratio and adiabatic sound speed are studied analytically after applying the boundary conditions.

Applying the boundary conditions from (11) to (14), we get the values of the arbitrary constants in terms of Schwarzschild parameters $u = \frac{GM}{c^2 R_b}$ and radius of the star $R_b$.

$$C = \frac{7(1-d)}{(7d-5)r_b^2} > 0 \quad for\, u \leq 0.244 \quad (32)$$

$$A = \frac{7.u.(1+Cr_b^2)^{\frac{5-S}{14}} - (5-S).Cr_b^2.d.(1+Cr_b^2)^{\frac{-9-S}{14}}}{(5+S)d.Cr_b^2.(1+Cr_b^2)^{\frac{-9+S}{14}} - 7.u.(1+Cr_b^2)^{\frac{5+S}{14}}} \quad (33)$$

$$B = \sqrt{\frac{(1+Cr_b^2)^{\frac{5-S}{14}} + A.(1+Cr_b^2)^{\frac{5+S}{14}}}{d}} \quad (34)$$

Where we 'd' given by

$$d = (1-2u)^{1/2}$$

Whose value lies between

$$0.715 < d < 1 \text{ for } Cr_b^2 > 0. \quad (35)$$

Surface density is given by

$$\rho_b R_b^2 = \frac{Cr_b^2[84 + 24Cr_b^2]}{49(1+Cr_b^2)^2} \quad (36)$$

Central red-shift is given by

$$Z_0 = \left[\frac{B^2}{1+A} - 1\right] \quad (37)$$

The surface red shift is given by

$$Z_b = [e^{-\frac{v_b}{2}} - 1] = d^{-1} - 1 \quad (38)$$

**Table 1**. *The effect of anisotropy on maximum radius and maximum mass of a star.*

| α | $u_{max}$ | $R_b$(km) | $M/M_\odot$ |
|---|---|---|---|
| 0.00 | 0.1728 | 12.68 | 1.47 |
| 0.01 | 0.1588 | 12.58 | 1.35 |
| 0.02 | 0.1302 | 12.11 | 1.06 |
| 0.03 | 0.0805 | 10.34 | 0.56 |
| 0.04 | 0.0050 | 2.82 | 0.01 |

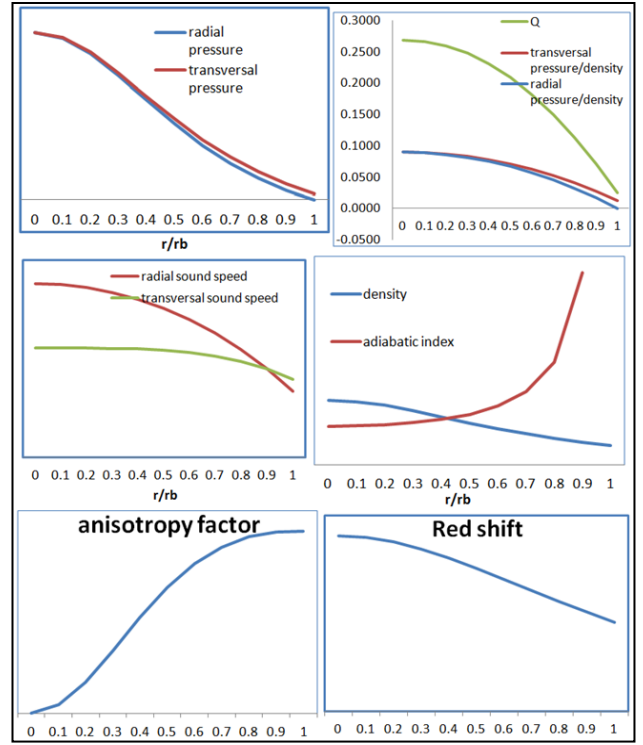

**Fig 1**. *The variation of* $P$, $\rho$, $Q$, $\frac{p}{\rho c^2}$, $Z$, $\left(\frac{dp}{d\rho}\right)$, $\gamma$, $\Delta$ *etc from centre to surface for α = 0.01 are shown in the following graphs.*

## 7. Discussions and Conclusions

From table 2 and figure 1 it has been observed that the physical quantities

( $P_r$, $P_\perp$, $\frac{P_r}{\rho}$, $\frac{P_\perp}{\rho}$, $\frac{dp_r}{d\rho}$, $\frac{dp_\perp}{d\rho}$, $z$, $Q$ ) are positive at the centre and within the limit of realistic state equation and monotonically decreasing while the quantities γ and Δ are increasing for all values of u satisfying the inequality 0.1728 > u >0. Thus, the solution is well behaved for all values 0 < u < 0.1728 for α lying in the range 0 to 0.04.

Table 1 shows that with the increase in the value of anisotropy in the fluid ball the mass decreases. This is because of diversion of pressure away from radial direction.

We present a model of super dense quark star based on the particular solution discussed above by assuming surface density $\rho_b = 2 \times 10^{14}\, g/cm^3$ the resulting well behaved solution has a maximum mass M = 1.35 $M_\oplus$ and radius R= 12.58 km.

*Table 2.* *The march of pressure, density, trace of energy stress tensor to density, square of adiabatic speed of sound within the ball for $\alpha = 0.01$ for which $u_{max}=0.1588$*

| $\frac{r}{r_b}$ | $p_r r_b^2$ | $p_\perp r_b^2$ | $\rho r_b^2$ | Q | $\frac{dp_r}{d\rho}$ | $\frac{dp_\perp}{d\rho}$ |
|---|---|---|---|---|---|---|
| 0 | 0.231149 | 0.231149 | 2.581626 | 0.26861 | 0.130053 | 0.1218862 |
| 0.1 | 0.223839 | 0.224296 | 2.525393 | 0.26627 | 0.129926 | 0.1218679 |
| 0.2 | 0.203583 | 0.205272 | 2.369259 | 0.25921 | 0.129546 | 0.1218107 |
| 0.3 | 0.174576 | 0.177944 | 2.144815 | 0.24732 | 0.128912 | 0.1217058 |
| 0.4 | 0.141735 | 0.146856 | 1.889215 | 0.23049 | 0.128021 | 0.1215363 |
| 0.5 | 0.109135 | 0.115804 | 1.633464 | 0.20860 | 0.126864 | 0.1212753 |
| 0.6 | 0.079348 | 0.087217 | 1.397399 | 0.18161 | 0.125425 | 0.1208874 |
| 0.7 | 0.053554 | 0.062245 | 1.190392 | 0.14957 | 0.123687 | 0.1203315 |
| 0.8 | 0.031973 | 0.041146 | 1.014533 | 0.11263 | 0.121632 | 0.1195654 |
| 0.9 | 0.014298 | 0.023677 | 0.867854 | 0.07104 | 0.119244 | 0.1185503 |
| 1 | 0.000000 | 0.009380 | 0.746636 | 0.02512 | 0.116513 | 0.1172536 |

## References

[1] Bowers, R.L., Liang, E.P.T.: Astrophys. J. 188, 657 (1974).

[2] Canuto, V., Lodenquai, J.: Phys. Rev. C 12, 2033 (1975).

[3] Chaisi, M., Maharaj, S.D.: Pramana journal of physics 66(3), 609 (2006).

[4] Das, B., Ray, P.C., Radinschi, I., Rahaman, F., Ray, S.: Int. J. Mod. Phys. D 20, 1675 (2011).

[5] Delgaty, M.S.R., Lake, K.: Comput. Phys. Commun. 115, 395 (1998).

[6] Esculpi,M.,Malaver,M.,Aloma,E.,:General Relativity and Gravitation ,39,633-652 (2007)

[7] Herrera, L., Ponce de Leon, J.: J. Math. Phys. 26, 2302 (1985).

[8] Herrera, L., Santos, N.O.: Phys. Rep. 286, 53 (1997).

[9] Herrera, L., Ospino, J., Prisco, A.D.: Phys. Rev. D 77, 027502 (2008).

[10] Herrera, L., Prisco, A.D., Ospino, J., Fuenmayor, E.: J. Math. Phys. 42, 2129 (2001).

[11] Maharaj, S.D., Takisa, P.M.: Gen. Relativ Gravit. 44, 1419 (2012).

[12] Maharaj, S.D., Sunzu, J.M., Ray, S.: Eur. Phys. J. Plus 129 (2014).

[13] Mak, M.K., Harko, T.: Chin. J. Astron. Astrophys. 3, 248 (2002).

[14] Mak, M.K., Harko, T.: Proc. R. Soc. Lond. A 459, 393 (2003).

[15] Mak, M.K., Dobson, P.N., Harko, T.: Int. J. Mod. Phys. D 11, 207 (2002).

[16] Malaver, M.: World Applied Programming., 3, 309 (2013).

[17] Malaver, M.: Open Science J. Mod. Phys. 1, 6 (2014)

[18] Maurya, S.K., Gupta, Y.K.: Phys. Scr. 86, 025009 (2012).

[19] Maurya, S.K., Gupta, Y.K.: Astrophys Space Sci 344, 243 (2013).

[20] Maurya, S.K., Gupta, Y.K.: Astrophys. Space Sci. 86, 025009 (2014).

[21] Pant. N,Pradhan,N.,Murad H.M.: Astrophys. Space Sci. (Accepted) 2014a,

[22] Pant.N., Pradhan,N., Ksh Newton Singh.: Journal of Gravity.doi.org/10.1155/2014/380320 2014b

[23] Ruderman, R.: A. Rev. Astr. Astrophys. 10, 427 (1972)